\documentclass{article}
\usepackage[utf8]{inputenc}
\usepackage[T1]{fontenc}

\usepackage{authblk}
\usepackage{amsmath}
\usepackage{graphicx}
\usepackage{verbatim}
\usepackage[subrefformat=parens]{subcaption}
\usepackage{siunitx}
\usepackage{enumitem}
\usepackage{bm}
\usepackage{float}
\usepackage[colorlinks=true,citecolor=blue,linkcolor=red]{hyperref}
\usepackage[ruled,vlined]{algorithm2e}
\usepackage{cleveref}
\usepackage{endnotes}

\let\footnote=\endnote

\usepackage[]{natbib}
\graphicspath{ {images/} }

\usepackage[normalem]{ulem}
\useunder{\uline}{\ul}{}

\title{A Map of Science in Wikipedia}
\author[1]{Puyu Yang}
\author[1]{Giovanni Colavizza\thanks{Please address correspondence to \url{p.yang2@uva.nl} and \url{g.colavizza@uva.nl}.}}
\affil[1]{Institute for Logic, Language and Computation (ILLC) \protect\\ University of Amsterdam, The Netherlands.}
\date{}
\begin{document}

\maketitle

\begin{abstract}
    In recent decades, the rapid growth of Internet adoption is offering opportunities for convenient and inexpensive access to scientific information. Wikipedia, one of the largest encyclopedias worldwide, has become a reference in this respect, and has attracted widespread attention from scholars. However, a clear understanding of the scientific sources underpinning Wikipedia's contents remains elusive. In this work, we rely on an open dataset of citations from Wikipedia to map the relationship between Wikipedia articles and scientific journal articles. We find that most journal articles cited from Wikipedia belong to STEM fields, in particular biology and medicine ($47.6$\% of citations; $46.1$\% of cited articles). Furthermore, Wikipedia's biographies play an important role in connecting STEM fields with the humanities, especially history. These results contribute to our understanding of Wikipedia's reliance on scientific sources, and its role as knowledge broker to the public.
\end{abstract}

\section*{Introduction}

Wikipedia is the largest, free and collaborative encyclopedia to date. Its importance cannot be overstated as Wikipedia provides reliable access to information worldwide. Wikipedia editors use primary and secondary sources in support of the statements they make in Wikipedia. These sources are quoted, cited and added to the list of references in any Wikipedia article. While previous work has focused on the contents of Wikipedia and its collaborative editing, it is only recently that scholars have begun to systematically investigate Wikipedia's sources.  

Developing a better understanding of the scientific sources Wikipedia relies on is important to assess its coverage, reliability and representation of human knowledge, including in view of informing its future development. Expanding upon previous work~\citep{teplitskiy2017amplifying,torres2019mapping,colavizza_covid-19_2020,arroyo2020science}, we pose here the following two descriptive research questions (RQs):
\begin{enumerate}
    \item RQ1: Which scientific sources are cited from Wikipedia and what are their characteristics?
    \item RQ2: Which areas of Wikipedia rely on scientific sources and how do they relate?
\end{enumerate}

In order to contribute towards answering them, in the present contribution we focus on citations given from Wikipedia articles to scientific journal articles. We rely on a recently published dataset as a source of such citation data~\citep{singh_wikipedia_2021}. Firstly, we provide a descriptive overview of the journal articles cited from Wikipedia (RQ1); secondly, we make use of network analysis and study the bibliographic coupling network of Wikipedia articles, respectively the co-citation network of journal articles (RQ2).

\section*{Previous work}

\paragraph{References in Wikipedia}

Wikipedia is a tertiary source which strives to provide reliable contents in a neutral way~\citep{10.1002/asi.23172}. To this end, Wikipedia editors follow established standards, guidelines and workflows to expand Wikipedia articles, and add references to them~\citep{kaffee2021references}. The bulk of the referencing activity seems to occur when an article has reached a certain level of maturity and number of edits. Furthermore, references ``also tend to be contributed by editors who have contributed more frequently and more substantially to an article, suggesting that a subset of more qualified or committed editors may exist for each article''~\citep{10.1145/2462932.2462943}. Despite this approach, Wikipedia's contents vary greatly in quality, including across languages~\citep{roy2021information}.

The automatic improvement of Wikipedia's contents is an area of active research. While bots already patrol and improve the quality of Wikipedia's references~\citep{zagovora2020updated}, recent work also focused on automatically flagging sentences in need for a citation~\citep{10.1145/3308558.3313618} and on assessing a source's reliability~\citep{lewoniewski2020modeling}.

\paragraph{Using Wikipedia}

Given its broad scope in contents, the usage of Wikipedia varies greatly too: ``for instance, we observe long and fast-paced page sequences across topics for users who are bored or exploring randomly, whereas those using Wikipedia for work or school spend more time on individual articles focused on topics such as science''\citep{10.1145/3038912.3052716}. The usage of Wikipedia is even more significant when considering countries with varied languages and socio-economic characteristics~\citep{10.1145/3289600.3291021}. Wikipedia also fulfills a specific role within the broader Web: it serves as a stepping stone between search engines and third-party websites~\citep{10.1145/3442381.3450136}. The complementarity of Wikipedia and search engines, such as Google, is particularly significant for scientific information seeking~\citep{10.1002/asi.23172}.

Previous studies have explored the use and usability of Wikipedia as a source of biomedical information. A randomized controlled trial focused on placing evidence of the effects of treatments (in this case, for schizophrenia) within Wikipedia pages finds no significant changes in the full-text accesses of the treated pages, but an effect in their altmetric scores~\citep{adams2020adding}. The readability of the most viewed Wikipedia articles on diseases is of varying quality, with many articles still too difficult to read for a general readership~\citep{brezar2019readability}. More generally, while Wikipedia is a prominent health information source in terms of views and visibility (e.g., often top in Google searches), the study of its impact in this respect is still too limited to draw any general conclusion~\citep{smith2020situating}.
    
A distinct set of recent studies explored the use of references in Wikipedia, and in particular references to external sources. Work on WikiProject Medicine shows that its readers appear to use links to external sources to verify and authorize Wikipedia content, rather than to examine the sources themselves~\citep{maggio2020meta}. A Wikipedia-wide study of engagement with external references found that users click on them only very rarely (once for every 300 page views, on average)~\citep{piccardi2020quantifying}. Crucially, users more often look for more when reading lower quality and shorter articles, which possibly do not contain what they seek~\citep{piccardi2021largescale}. These findings further underline the importance of providing high-quality contents within Wikipedia itself.  
    
\paragraph{Science and Wikipedia}

Wikipedia strives to convey information grounded in scientific results. It thus provides visibility to scientific research, and is considered an altmetric source in this respect~\citep{sugimoto2017scholarly}, yet the influence goes both ways. In fact, previous work has established that being cited from Wikipedia can increase the citation impact of an article~\citep{thompson2018science}. Understanding and monitoring which scientific results underpin Wikipedia's contents, and why, is therefore of critical importance.

Recent work is gradually improving our understanding of the matter. The open release of datasets of citations from Wikipedia to its sources is helping in broadening access to the essential data for tackling the question at hand~\citep{halfaker_2018,zagovora2020updated,singh_wikipedia_2021}. Furthermore, several previous studies have been able to rely on altmetric data. Some trends clearly emerge from this literature. Journal articles cited from Wikipedia are more likely than average published in high-impact journals (e.g., by impact factor), and in open access~\citep{nielsen2007scientific,teplitskiy2017amplifying}. Articles cited from Wikipedia are `uncited' and untested by subsequent studies in rates proportional to the rest of the scientific literature, nevertheless they also receive a higher rate of supporting citations~\citep{nicholson2021measuring}.

Wikipedia's capacity to rapidly and reliably integrate novel scientific results to respond to ongoing public events or crises has also been assessed. The COVID-19 pandemic provides a recent example~\citep{colavizza_covid-19_2020}. Most notably, the areas of Wikipedia where the editors are highly organized and include domain experts, for example several WikiProjects, appear to fare better in this respect. Indeed, the scope of expert involvement in editing Wikipedia is substantial. A recent study found that approximately 10\%--30\% of Wikipedia’s contributors have substantial subject-matter expertise in the topics that they edit~\citep{yarovoy2020assessing}.

Recent results by \cite{arroyo2020science}, extending previous work by the same team~\citep{torres2019mapping}, directly relate to our study. The authors perform a co-citation analysis of Wikipedia's sources, relying on a dataset of ``847,512 references made by 193,802 Wikipedia articles to 598,746 scientific articles belonging to 14,149 journals indexed in Scopus.'' They use Altmetrics data to retrieve Wikipedia citations to journal articles. Their study the co-citation network of journals and Scopus main field categories, as referenced by Wikipedia. The most significant scientific domains cited by Wikipedia include Medicine, Biochemistry, Genetics, and Molecular Biology. They confirm that the most important journals are multidisciplinary, and include prominent venues such as Nature, Science, PNAS, and that articles from high-impact factor journals are more likely to be cited from Wikipedia. Lastly, they find that only 13.44\% of Wikipedia citations are to open access journals. Directly expanding upon \cite{arroyo2020science}, we use here a larger and more recent dataset, considering the more granular level of analysis of journal articles, and we deepen the analysis by comparing the bibliographic coupling network of Wikipedia articles with the co-citation network of journal articles.
    
In conclusion, it is also worth noting that understanding how Wikipedia is structured according to how it uses scientific publications complements work relying on its internal link network. For example, previous results have found that different scientific domains possess distinct internal link network organizations, with modular structures for Biology and Medicine, but a sparse structure for Mathematics and a dense core for Physics~\citep{SILVA2011431}.

\section*{Data and methods}

We assemble data from a variety of sources to perform our study.

\paragraph{Wikipedia Citations}

We use \textit{Wikipedia Citations} as our main dataset~\citep{singh_wikipedia_2021}. It consists of more than 29M citations extracted from the over 6M articles composing the English Wikipedia as of May 2020. In \textit{Wikipedia Citations}, each citation is automatically classified as being to a book, journal article or Web content. Approximately 2.5M citations are classified as to a journal article, of which 1,705,085 are equipped with a DOI, either from Wikipedia itself or retrieved from Crossref. These citations to journal articles come from 405,358 distinct Wikipedia article pages and refer to 1,157,571 distinct DOIs. Citations to journal articles clearly comprise a relatively limited share of all citations contained in Wikipedia, and thus likely serve a specific purpose~\citep{singh_wikipedia_2021}. A large share of these citations are given to articles published over the past 20 years, and the most cited journals include Nature, Science, the Journal of Biological Chemistry and PNAS. We use this set of citations in what follows.

\paragraph{ORES Topics, WikiProjects, Dimensions}

In order to perform our analysis, we enrich the \textit{Wikipedia Citations} dataset with other sources of information. Firstly, we equip Wikipedia articles with information about their topics and the WikiProject they belong to, if any. The topics of Wikipedia articles are retrieved using the ORES Web service\footnote{We made use of the second API at the following address: \url{https://wiki-topic.toolforge.org/\#lang-agnostic-model}.}, which exposes a topic model of Wikipedia trained using Language-Agnostic Topic Classification (LATC)~\citep{johnson2021language} and assigns each Wikipedia article to a taxonomy rooted into the four categories of: Geography, Culture, History and Society, and STEM.\footnote{\url{https://www.mediawiki.org/wiki/ORES/Articletopic\#Taxonomy}.} Through the ORES API, we could extract topics for Wikipedia articles covering 99.7\% of the citations in \textit{Wikipedia Citations}.

Further, we equip Wikipedia articles with information on their WikiProject. ``A WikiProject is a group of contributors who want to work together as a team to improve Wikipedia''\footnote{\url{https://en.wikipedia.org/wiki/Wikipedia:WikiProject}.}, for example to focus on a specific topic area such as WikiProject Mathematics or WikiProject India, or curate a specific aspect of the encyclopedia, for instance WikiProject Disambiguation. The English Wikipedia currently includes over 2,000 WikiProjects. Using public data~\citep{Johnson2020}, we could equip with WikiProjects information Wikipedia articles comprising 96.3\% of the citations in \textit{Wikipedia Citations}.

Finally, we use Dimensions~\citep{herzog_dimensions_2020} to retrieve metadata for all the journal articles with a DOI cited from Wikipedia. While no ideal single bibliographic data source exists yet, Dimensions provides for broad source coverage and relies in substantial part on the open Crossref repository, making it a meaningful choice for our study~\citep{10.1162/qss_a_00112}. What is more, we are further interested in an article's Field of Research (FOR) classification.\footnote{The Fields of Research classification follows the research areas defined in the Australian and New Zealand Standard Research Classification (ANZSRC). See: \url{https://app.dimensions.ai/browse/categories/publication/for}.} The Fields of Research are organized into hierarchies, with divisions (top, largest), groups and fields (bottom, smallest). Dimensions exposes divisions as \textit{major fields}, and groups as \textit{minor fields}; we adopt this naming convention in what follows. By querying the Dimensions' API we were able to match 96\% of all the unique DOIs from \textit{Wikipedia Citations}. All these data were retrieved in June 2021.

\subsection*{Methods}

For our study we make a comparison between two undirected networks in turn extracted from the directed citation network of Wikipedia to journal articles: the co-citation network among scientific journal articles, and the bibliographic coupling network among Wikipedia articles. Using the conceptual framework of \cite{costas_heterogeneous_2021}, we can say that the co-citation network is made of co-Wikipedia citations whereby two journal articles are cited by the same Wikipedia article(s), and the bibliographic coupling network is composed of Wikipedia articles connected when they cite the same journal articles. Both networks are clustered using the Leiden algorithm~\citep{traag2019louvain}, and their modular structures are compared. In this way, we aim to clarify which areas of Wikipedia rely on scientific sources. What is more, we seek to show how similar clusters of Wikipedia articles (bibliographic coupling) and similar clusters of co-cited scientific articles (co-citation) are internally related (RQ2).

\paragraph{Citation networks}

While constructing the co-citation network, we remove nodes (journal articles) that are cited only once from Wikipedia, as they would be isolated in the co-citation network since they are never co-cited. This results in a network of 1,050,686 nodes (91\% of 1,157,571) and 17,916,861 edges. Similarly, we remove nodes (Wikipedia articles) that cite only one journal article, as they would be isolated in the bibliographic network. This gives a network of 257,452 nodes (64\% of 405,358) and 27,473,262 edges. This bibliographic coupling network is thus not only denser, but it also does not include a higher share of isolated nodes than the co-citation network. Nodes in both networks are equipped with relevant metadata: WikiProject and ORES topics for the bibliographic coupling network, Fields of Research for the co-citation network. Furthermore, in both networks every edge is weighted according to how many times any two nodes are cited together (co-citation) or jointly cite the same items (bibliographic coupling).

\paragraph{Network clustering}

\begin{figure}[H]
\begin{subfigure}{.5\textwidth}
  \centering
  \includegraphics[width=.99\linewidth]{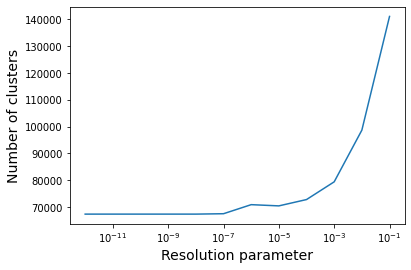}
  \caption{Co-citation.}
  \label{fig:respar_cocit}
\end{subfigure}%
\begin{subfigure}{.5\textwidth}
  \centering
  \includegraphics[width=.99\linewidth]{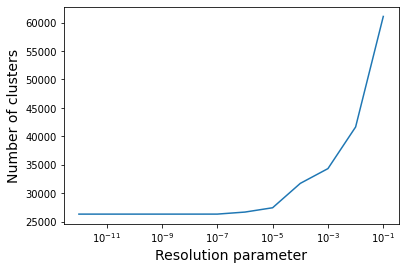}
  \caption{Bibliographic coupling.}
  \label{fig:respar_bibc}
\end{subfigure}
\caption{Number of clusters at varying values of the resolution parameter.}
\label{fig:respar}
\end{figure}

In order to equip our networks with a clustering solution, we use the Leiden algorithm~\citep{traag2019louvain}, a popular choice that is fast and provides specific guarantees over the resulting clustering.\footnote{For our analyses we relied on \textit{igraph} 0.9.6 and \textit{leidenalg} 0.8.3.} The Leiden algorithm uses a resolution (hyper)parameter to control the balance between the number of clusters and their size. A higher resolution parameter leads to more, smaller clusters. We find a reasonable value for the resolution parameter empirically, by inspecting the number of clusters at varying values of the parameter. Figure~\ref{fig:respar_cocit} shows the number of clusters at the varying size of the resolution parameter for the co-citation network, and Figure~\ref{fig:respar_bibc} does the same for the bibliographic coupling network. We avoid choosing extreme values, and settle for reasonable elbows which can be found, in both cases, at a value of the resolution parameter of $1e-4$.

\paragraph{Supernetworks}

\begin{figure}[H]
\begin{subfigure}{.5\textwidth}
  \centering
  \includegraphics[width=.99\linewidth]{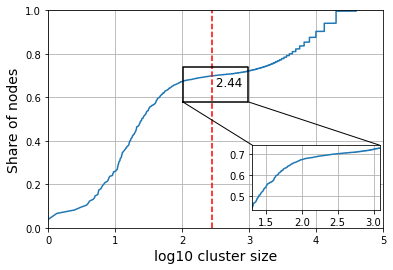}
  \caption{Co-citation.}
  \label{fig:thres_cocit}
\end{subfigure}%
\begin{subfigure}{.5\textwidth}
  \centering
  \includegraphics[width=.99\linewidth]{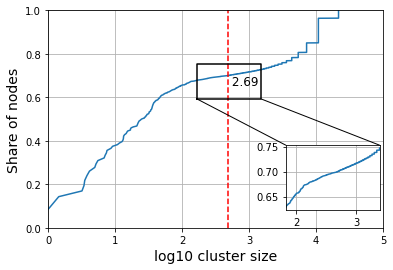}
  \caption{Bibliographic coupling.}
  \label{fig:thres_bibc}
\end{subfigure}
\caption{Cumulative share of nodes included in the supernetworks, per cluster size threshold. Clusters are ordered from largest (left) to smallest (right).}
\label{fig:thres}
\end{figure}

In order to further coarsen our networks, we also construct `supernetworks' of clusters. In a supernetwork, each node is a cluster of nodes from an underlying network. A supernetwork is therefore a network of clusters. A supernetwork is often easier to visualize and inspect, as it contains a much smaller number of nodes than the original network. We construct supernetworks by weighting each node with the number of nodes that a cluster contains, and by weighting each edge by summing the weights of the edges between the nodes of any two clusters. The use of aggregated metadata from the nodes of a cluster is detailed below, when discussing results. In order to identify and focus on the largest and most representative clusters, we further trim the supernetworks at a given cluster size threshold. We do so empirically, by inspecting the cumulative share of nodes that would be included in a supernetwork, at varying cluster size thresholds. Results are shown in Figure~\ref{fig:thres_cocit} (co-citation) and Figure~\ref{fig:thres_bibc} (bibliographic coupling). In both cases, reasonable cutoffs can be found at elbows corresponding to a cumulative share of included nodes of $.7$. This cutoff corresponds in turn to cluster sizes of 98 (co-citation) and 48 (bibliographic coupling), below which clusters are removed from the supernetworks.

\section*{Results}

We organise our results into three parts, addressing RQ1 in the first, and RQ2 in the second and third sections. First, we provide an overview of the scientific sources cited from Wikipedia. Next we analyse how Wikipedia articles are organised accordingly using bibliographic coupling networks. Lastly, we analyse how journal articles cited from Wikipedia are organised in turn, using co-citation networks. 

\subsection*{Science in Wikipedia}

\begin{table}[H]
\begin{tabular}{lll}
Field of Research (major field)                                       & Citations & Journal articles \\
06 Biological Sciences                          & 440,029.8(25.8\%)   & 255,382.0(22.1\%)  \\
11 Medical and Health Sciences                  & 371,304.7(21.8\%)   & 278,289.3(24.0\%)  \\
04 Earth Sciences                               & 69,835.7(4.1\%)     & 42,865.6(3.7\%)    \\
02 Physical Sciences                            & 69,678.3(4.1\%)     & 42,405.1(3.7\%)    \\
03 Chemical Sciences                            & 67,931.0(4.0\%)     & 53,396.7(4.6\%)    \\
17 Psychology and Cognitive Sciences            & 64,092.0(3.8\%)     & 48,072.9(4.2\%)    \\
21 History and Archaeology                      & 63,292.9(3.7\%)     & 40,855.7(3.5\%)    \\
16 Studies in Human Society                     & 57,414.7(3.4\%)     & 40,640.1(3.5\%)    \\
09 Engineering                                  & 42,994.5(2.5\%)     & 34,626.0(3.0\%)    \\
01 Mathematical Sciences                        & 41,408.8(2.4\%)     & 31,755.2(2.7\%)    \\
08 Information and Computing Sciences           & 33,286.9(2.0\%)       & 25,300.0(2.2\%)    \\
20 Language, Communication and Culture          & 31,969.4(1.9\%)     & 23,256.9(2.0\%)   \\
05 Environmental Sciences                       & 22,993.5(1.3\%)     & 15,069.7(1.3\%)    \\
22 Philosophy and Religious Studies             & 21,448.8(1.3\%)     & 15,654.1(1.4\%)    \\
14 Economics                                    & 182,30.9(1.1\%)     & 13,720.5(1.2\%)    \\
07 Agricultural and Veterinary Sciences         & 15,525.6(0.9\%)     & 12,146.2(1.0\%)     \\
15 Commerce, Management, Tourism and   Services & 12,691.6(0.7\%)     & 10,146.7(0.9\%)    \\
13 Education                                    & 10,534.9(0.6\%)     & 8,714.2(0.8\%)     \\
19 Studies in Creative Arts and Writing         & 10,304.9(0.6\%)     & 8,085.8(0.7\%)     \\
18 Law and Legal Studies                        & 9,971.6(0.6\%)      & 7,189.6(0.6\%)     \\
10 Technology                                   & 5,740.8(0.3\%)      & 4,433.4(0.4\%)     \\
12 Built Environment and Design                 & 2,886.5(0.2\%)      & 2,274.6(0.2\%)     \\
Missing                                         & 221,517.0(13.0\%)   & 143,291.0(12.4\%)  \\
Total                                           & 1,705,085(100\%)    & 1,157,571(100\%)  
\end{tabular}
\caption{Number of citations and of journal articles per Field of Research (major fields), using fractional counting.}\label{tab:table1}
\end{table}

\paragraph{Fields of Research} Table~\ref{tab:table1} shows the number of citations and the number of journal articles cited from Wikipedia per Dimension's major Field of Research. Given that journal articles (and, later on, Wikipedia articles) are our units of analysis, we use fractional counting to account for an article belonging to multiple categories~\citep{PERIANESRODRIGUEZ20161178}, unless otherwise specified. It can be seen that almost half of citations are given to articles in the fields of biological, medical and health sciences, which in turn make up about 46\% of all journal articles cited from Wikipedia. More in general, STEM fields make up for a large part of both citations and articles, while the most represented areas outside of STEM are history and sociology. Looking at minor Fields of Research, in Table~\ref{tab:table4} (only the top 10 are shown), we can appreciate how the most cited fields include genetics and cell biology, evolutionary biology and ecology in biology; clinical science and public health, psychology and neurosciences in medicine; history in the humanities. These fields capture over a fifth of all citations from Wikipedia.
    
\begin{table}[H]
\begin{tabular}{lll}
Field of Research (minor fields)                                & Fractional counting & Unique dois number \\
0604 Genetics                            & 76,299.8(4.5\%)      & 39,393.3(3.4\%)     \\
0601 Biochemistry and Cell   Biology     & 71,505.5(4.2\%)      & 45,098.7(3.9\%)     \\
1103 Clinical Sciences                   & 49,352.5(2.9\%)      & 38,700.8(3.3\%)     \\
1117 Public Health and Health   Services & 32,066.8(1.9\%)      & 23,282.4(2.0\%)     \\
1701 Psychology                          & 27,375.5(1.6\%)      & 20,623.2(1.8\%)     \\
1109 Neurosciences                       & 25,030.1(1.5\%)      & 19,117.9(1.7\%)     \\
2103 Historical Studies                  & 24,134.6(1.4\%)      & 15,873.0(1.4\%)     \\
0403 Geology                             & 23,114.6(1.4\%)      & 13,464.6(1.2\%)     \\
0602 Ecology                             & 23,033.6(1.4\%)      & 14,862.2(1.3\%)     \\
0603 Evolutionary Biology                & 18,246.4(1.1\%)      & 8,667.2(0.7\%)     
\end{tabular}
\caption{Number of citations and of journal articles per Field of Research (minor fields), using fractional counting.}\label{tab:table4}
\end{table}
    
\paragraph{Citation counts} We assess the distribution of (journal article) citations given to articles cited from Wikipedia, using data from Dimensions. There is a wide variation in citations counts, with many articles with no or few citations, and some with a high number. The maximum number of received citations is 214,886 and the minimum is 0, with a mean at 189 and a median at 33. When considering recent citations (over the past two years), the maximum is 34,845, the minimum is 0, with a mean at 36.4 and a median at 5. Furthermore, 60\% of articles cited from Wikipedia received fewer than 10 citations in the past two years. Articles cited fewer than 100 times account for 70\% of the total cited articles, and only about 3\% of articles are cited 1,000 times or more. It is worth noting that 10.4\% of journal articles are never cited and 22.5\% have no recent citations either, according to Dimensions. Finally, we note that 75,248 (4.4\%) articles were missing this information from Dimensions. 

\begin{comment}
\begin{figure}[H]
\centering
\includegraphics[width=\textwidth]{Figure3.png}
\caption{Number of (journal article) citations received by journal articles cited from Wikipedia. Data from Dimensions.}
\label{fig:citation_counts}
\end{figure}
\end{comment}

\paragraph{Most cited articles and journals} Among the top journal articles by a number of received citations (See Appendix, Section~\ref{sec:top_journal_articles}), only two are open access, six are authored by scholars based in the United States, and only one is published after 2000. The most cited article is ``Cleavage of Structural Proteins during the Assembly of the Head of Bacteriophage T4'', published in Nature in 1970, and cited 214,886 times according to Dimensions. Instead, the most cited article in terms of recent citations (past two years) is ``Deep Residual Learning for Image Recognition'', published at the 2016 Conference on Computer Vision and Pattern Recognition, with 34,845 recent citations. The top citing Wikipedia articles, instead, frequently include reference or survey articles. For example, the series ``this year in'' is often a source of highly citing Wikipedia articles (e.g., ``2018 in paleontology'' includes 580 citations to journal articles). In Table~\ref{tab:table9}, we list the most cited journals in Wikipedia. Nature, PNAS and Science top the list, confirming findings from previous work~\citep{arroyo2020science}.

\begin{table}[H]
\centering
\begin{tabular}{lr}
\centering
\textbf{Journal name}           & \textbf{Citations} \\
Nature                          & 37,287            \\
PNAS                            & 31,801            \\
Science                         & 26,903            \\
Journal of Biological Chemistry & 24,518            \\
PLOS ONE                        & 12,997            \\
Zootaxa                         & 10,006            \\
Cell                            & 9,318             \\
Genome Research                 & 8,961             \\
The Astrophysical Journal       & 8,882             \\
Astronomy \& Astrophysics       & 7,292            
\end{tabular}
\caption{Most cited journals}\label{tab:table9}
\end{table}
    
\begin{figure}[H]
\centering
\includegraphics[width=0.8\textwidth]{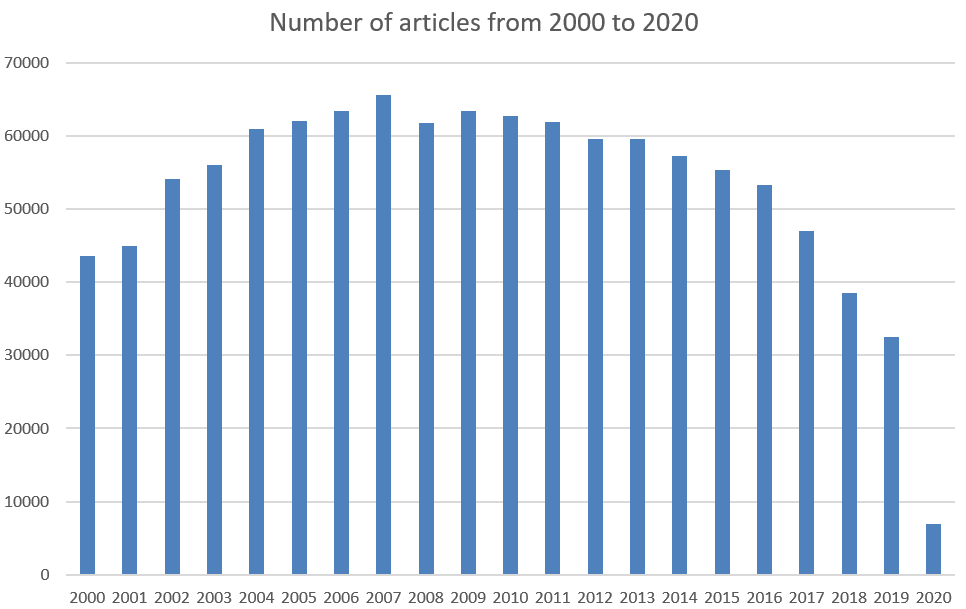}
\caption{Publication year of journal articles cited from Wikipedia, from 2000 to 2020.}
\label{fig:Number of articles from 2000 to 2020}
\end{figure}

\paragraph{Age of articles and open access} In Figure~\ref{fig:Number of articles from 2000 to 2020}, we show the distribution of the publication years for journal articles cited from Wikipedia. The trend is clear: most cited articles were published in the past two decades. We also assess the open access availability of these journal articles using Dimensions data. We find that 686,952 (41\%) articles are available in some form of open access, while a majority number of 942,885 (55\%) remain closed access, and for 75,248 (4\%) this information is missing. This constitutes a considerably higher fraction of cited OA articles than what previous work has found~\citep{arroyo2020science}.

\begin{figure}[H]
\centering
\includegraphics[width=0.8\textwidth]{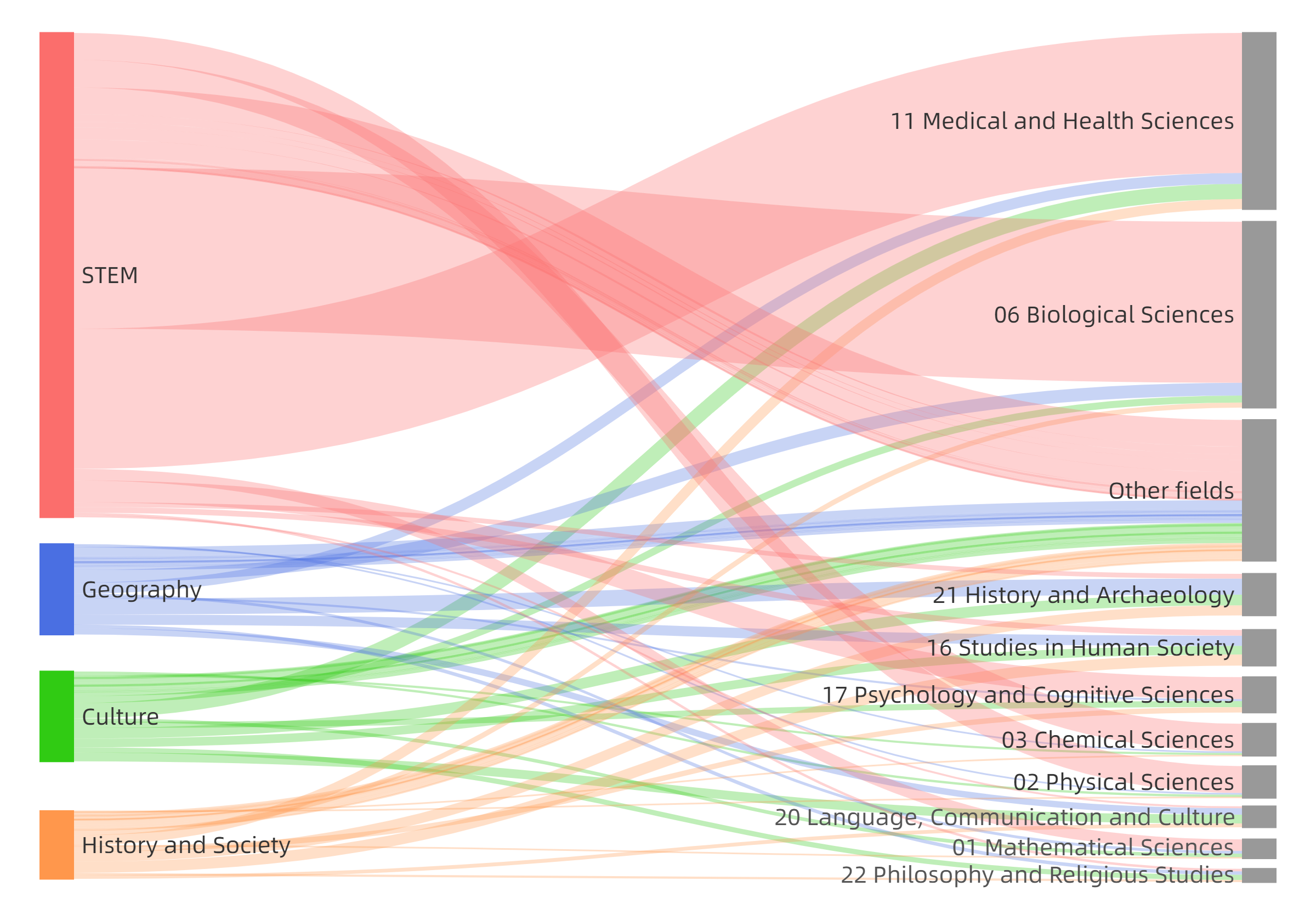}
\caption{Citation flow from ORES topics to major fields of research.}
\label{fig:River plot from Topics to fields}
\end{figure}

\begin{figure}[H]
\centering
\includegraphics[width=0.8\textwidth]{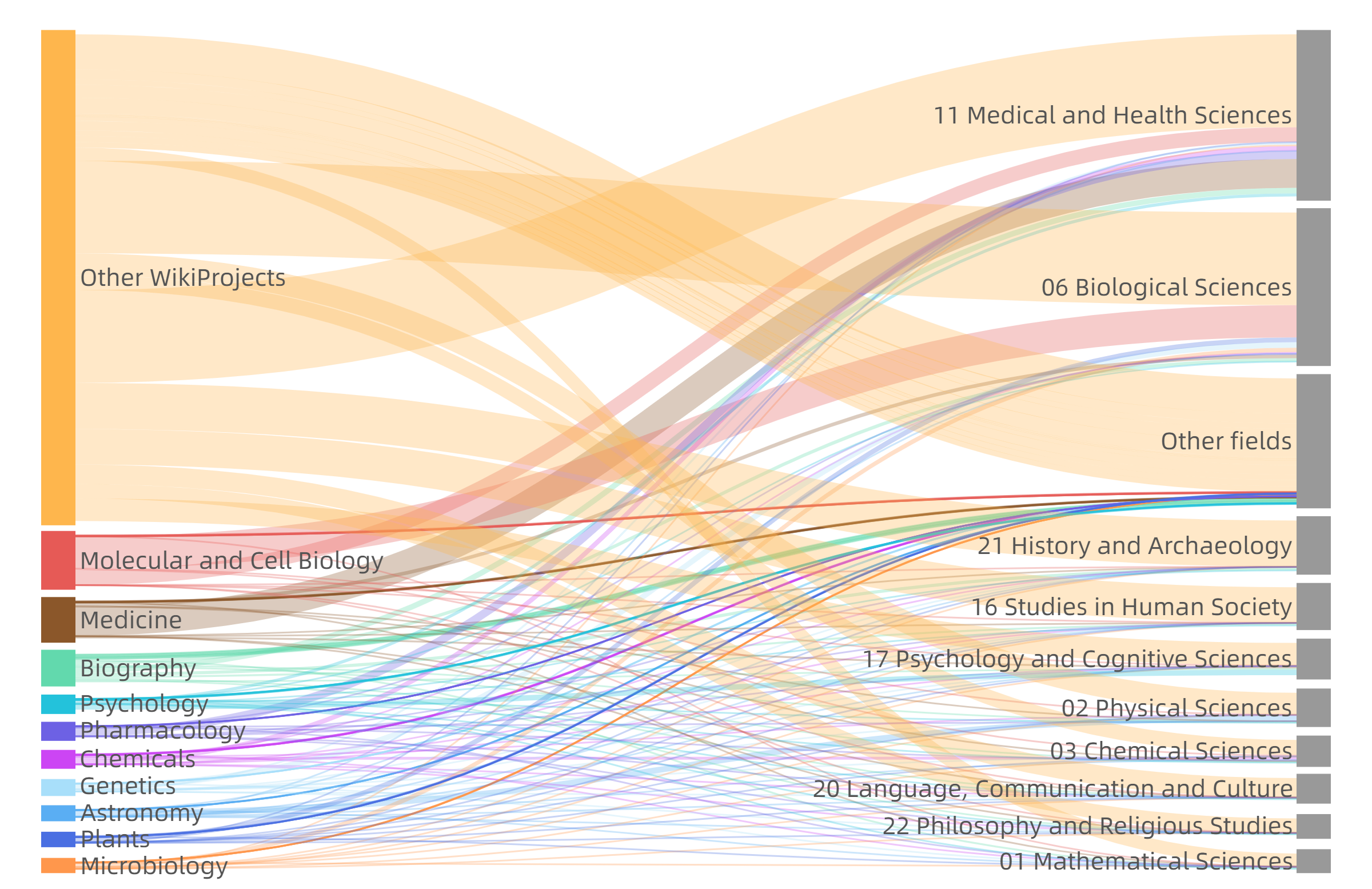}
\caption{Citation flow from the top-10 WikiProjects to major fields of research.}
\label{fig:River plot from top 10 wiki projects to fields}
\end{figure}

\paragraph{Citation flows} In conclusion of this first overview section, we analyse the flow of citations from Wikipedia to journal articles, on the one hand, grouping Wikipedia articles by ORES topics and WikiProjects, and on the other hand, grouping journal articles by major FOR categories. We show the river plots of citations from Wikipedia articles by ORES topics in Figure~\ref{fig:River plot from Topics to fields} and by (top-10) WikiProjects in Figure~\ref{fig:River plot from top 10 wiki projects to fields}. The flow of citations from STEM Wikipedia articles confirms the importance of the biological, medical and health sciences in Wikipedia, while other topics are more evenly distributed across fields of research. When considering WikiProjects, we only kept the top-10 by (fractional) number of Wikipedia articles, which have a strong STEM focus. Partially as a consequence, the flow of citations again favours the biological, medical and health sciences, but also distributes across other STEM fields of research. It is worth highlighting the role of the WikiProject Biography, which focuses on the biographies of notable persons. This project spans across fields of research, for example by covering the lives of prominent scientists, and connects them with historical fields in turn.

\subsection*{Wikipedia from a science perspective}
  
\begin{figure}[H]
\centering
\includegraphics[width=\textwidth]{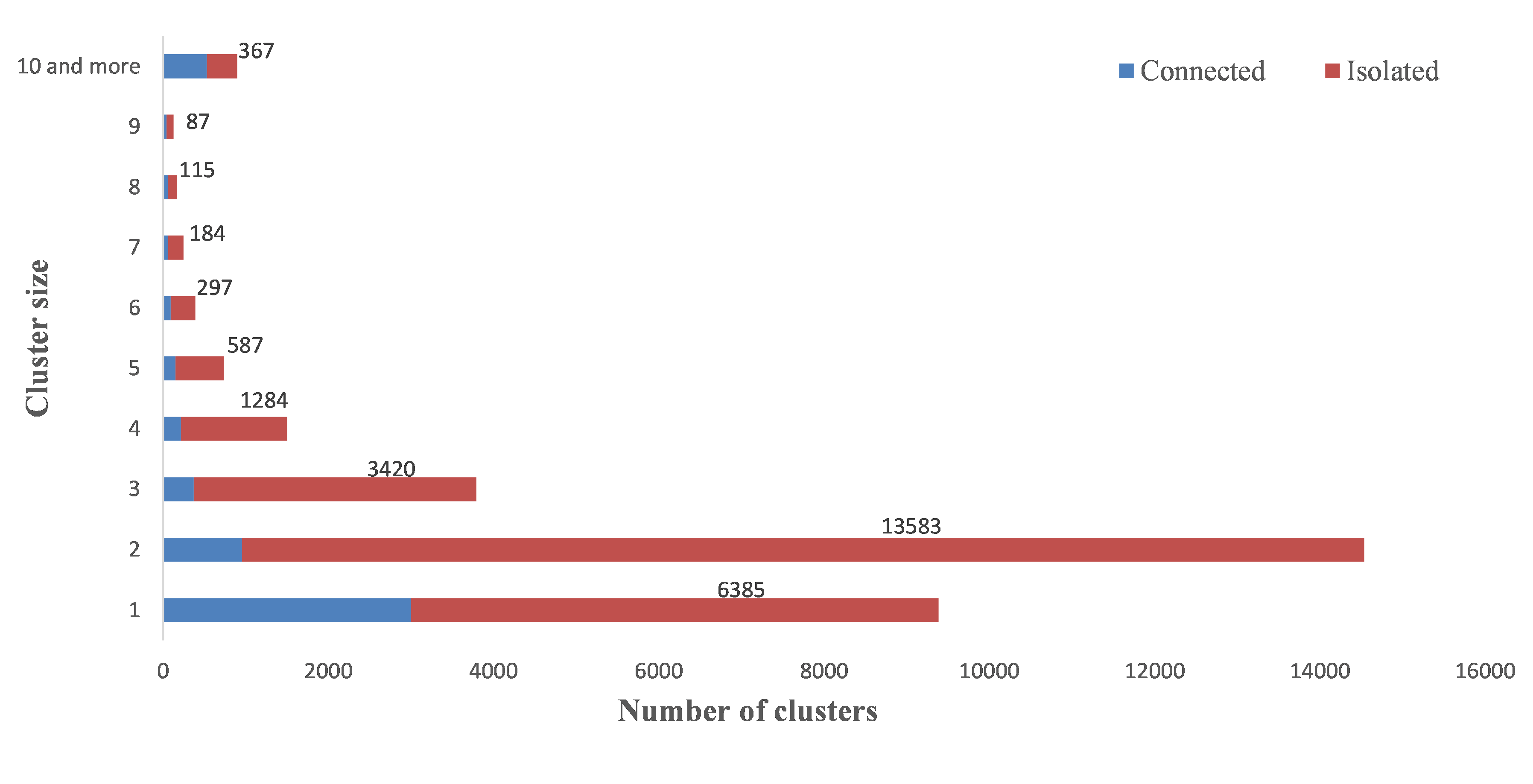}
\caption{Distribution of clusters by size in the bibliographic coupling network of Wikipedia articles.}
\label{fig:distribution_biblio}
\end{figure}

\paragraph{Bibliographic coupling network} The bibliographic coupling network of Wikipedia articles includes 257,452 nodes and 27,473,262 edges, with a density of 0.0008. The corresponding supernetwork has 31,642 nodes (clusters) and 9,158 edges, for a density of 0.000018. We further aggregate information about ORES topics and  WikiProjects to the supernetwork using fractional counting. In Figure~\ref{fig:distribution_biblio} we show the distribution of the cluster sizes of the underlying bibliographic coupling network, distinguishing between isolated and connected clusters. A large number of small, isolated clusters exists, while at higher cluster sizes the clusters become increasingly connected.
    
    \begin{figure}[H]
    \centering
    \includegraphics[width=\textwidth]{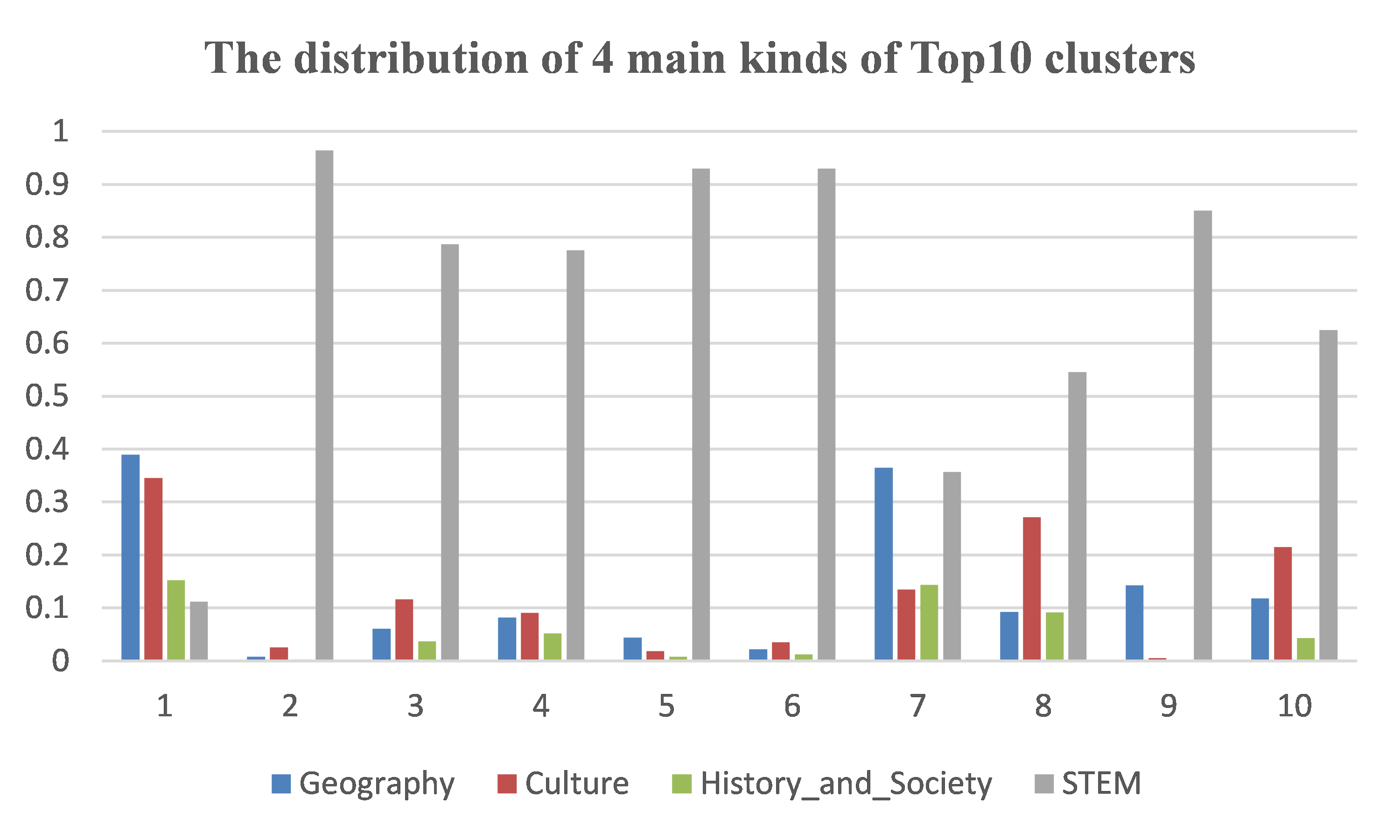}
    \caption{Distribution of ORES topics in the 10 largest clusters of the bibliographic coupling network of Wikipedia articles.}
    \label{fig:top10_topic_biblio}
    \end{figure}
    
\begin{figure}[H]
\centering
\includegraphics[width=\textwidth]{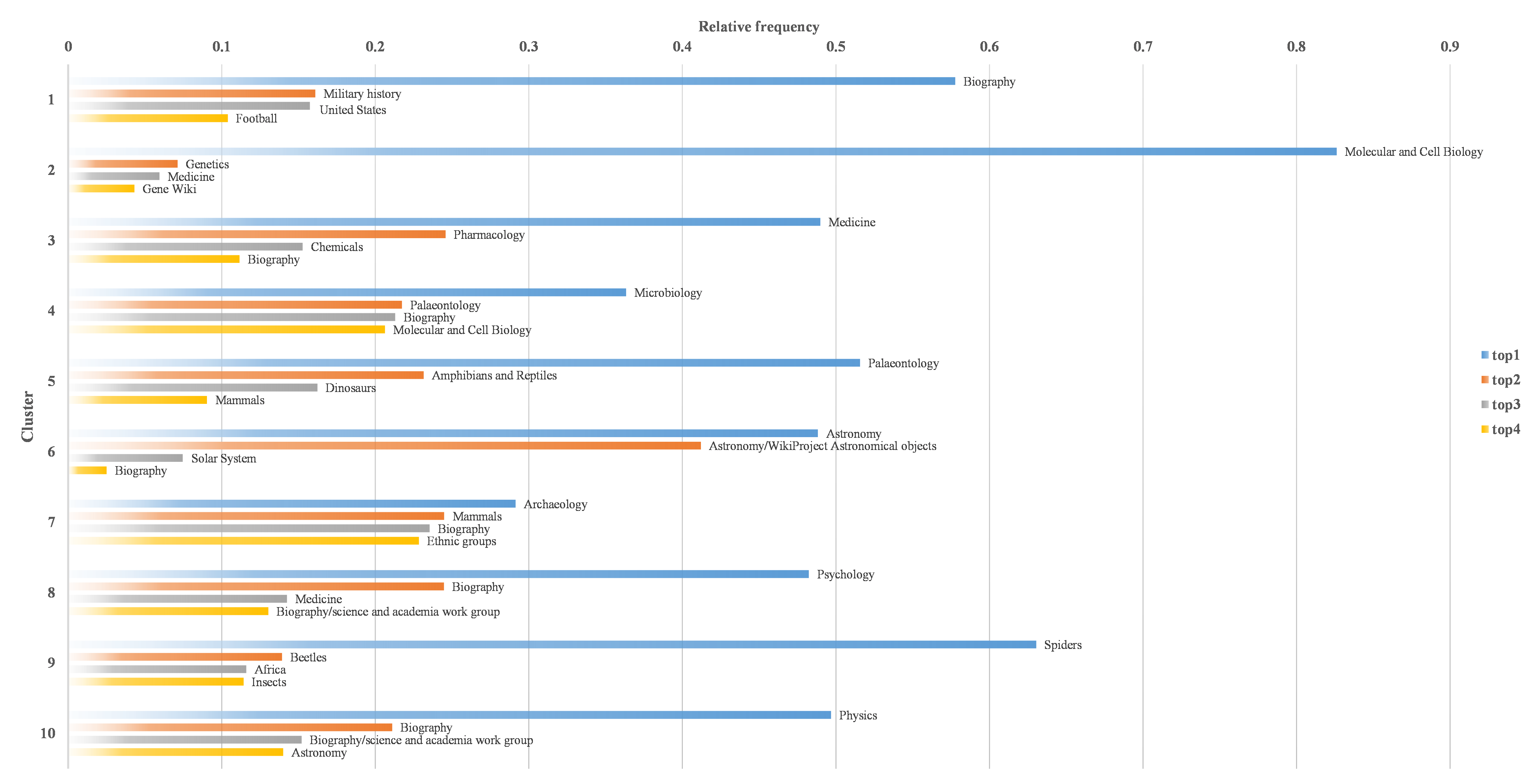}
\caption{Distribution of WikiProjects in the 10 largest clusters of the bibliographic coupling network of Wikipedia articles.}
\label{fig:top10_wp_biblio}
\end{figure}

\paragraph{Largest clusters} Next, we zoom-in the largest clusters of this network, showing their ORES topic distribution in Figure~\ref{fig:top10_topic_biblio}, and their distribution over WikiProjects in Figure~\ref{fig:top10_wp_biblio}. The combination of topics and projects is helpful in uncovering what a cluster is about. For example, the largest cluster is devoted to biographies and has a relatively balanced topic distribution across geography, history and society, culture. Importantly though, its top WikiProjects include military history, United States and football, suggesting a certain skew in this respect. Most clusters are, instead, specialized in specific STEM fields. For example, cluster 2 is focused on molecular and cell biology, and cluster 9 on entomology (spiders, in particular). The dominance of STEM in `journal-article-citing' Wikipedia is further confirmed. Exceptions include history and biographies, and archaeology/physical anthropology (cluster 7).

\begin{figure}[H]
\centering
\includegraphics[width=\textwidth]{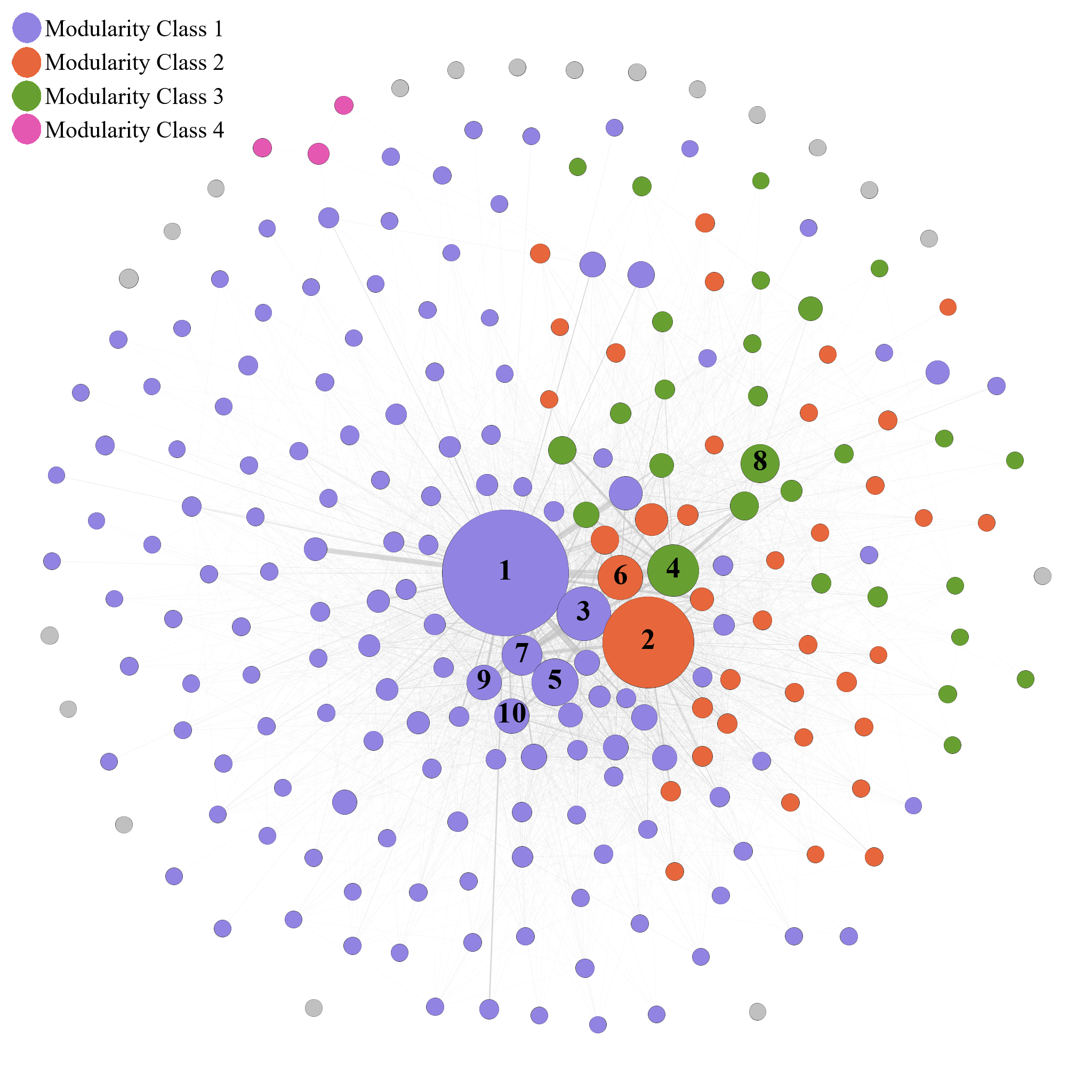}
\caption{Bibliographic coupling supernetwork coloured by modularity class.}
\label{fig:biblio_clustered_modularity}
\end{figure}

\begin{figure}[H]
\centering
\includegraphics[width=\textwidth]{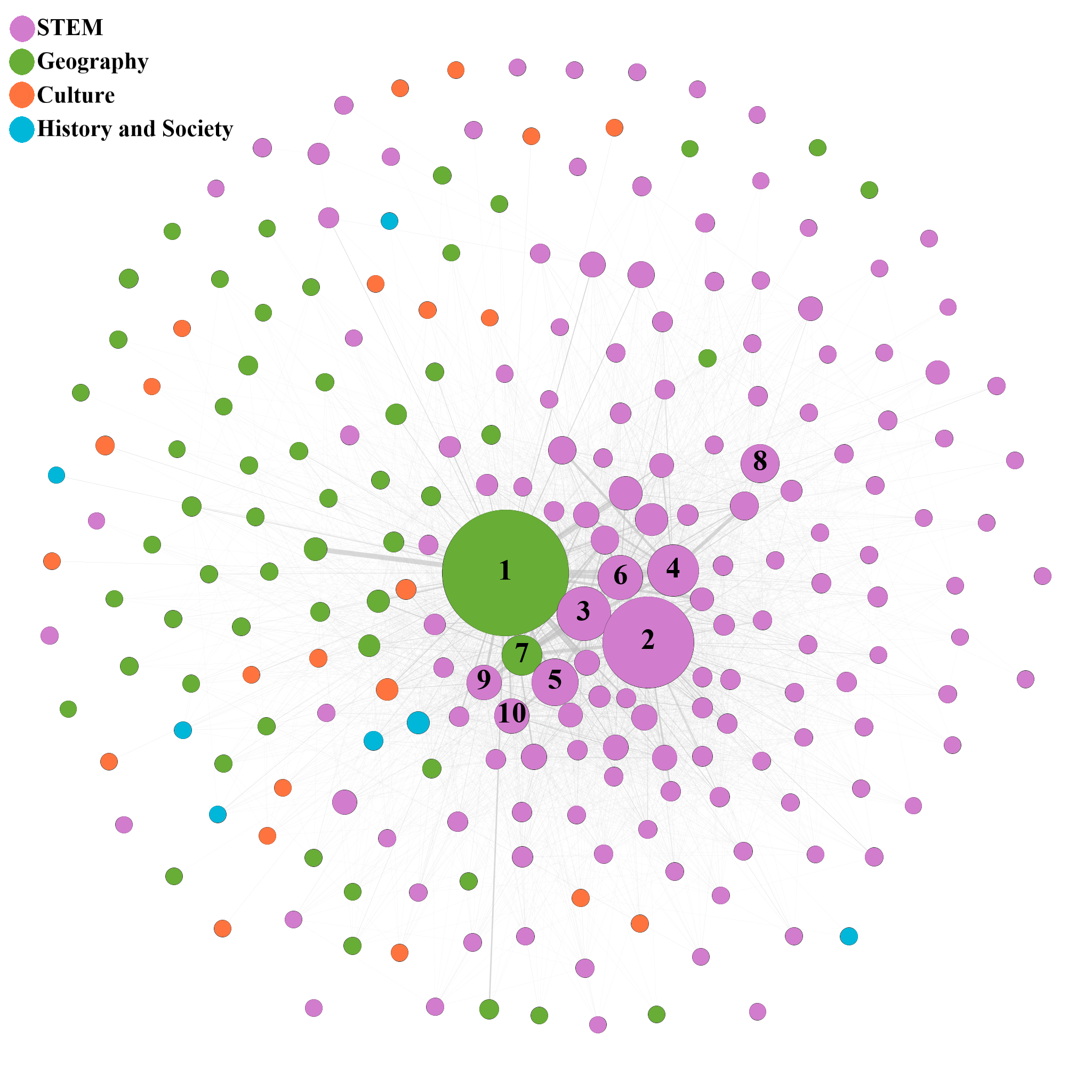}
\caption{Bibliographic coupling supernetwork coloured by top ORES topic within a node/cluster. Compare with Figure~\ref{fig:top10_topic_biblio}.}
\label{fig:biblio_clustered_topics}
\end{figure}

\begin{figure}[H]
\centering
\includegraphics[width=\textwidth]{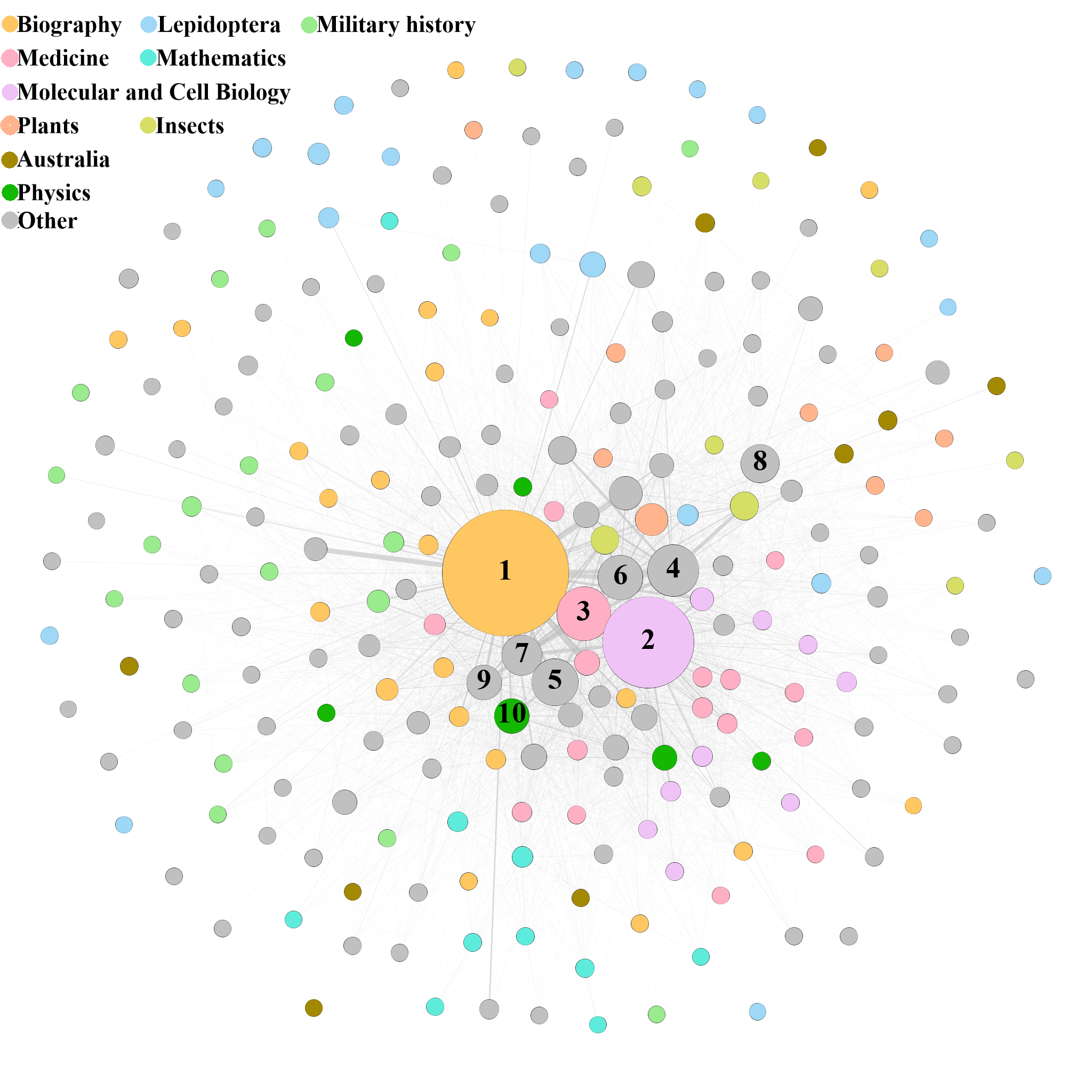}
\caption{Bibliographic coupling supernetwork coloured by top WikiProject within a node/cluster. Compare with figure~\ref{fig:top10_wp_biblio}.}
\label{fig:biblio_clustered_wp}
\end{figure}

\paragraph{Network visualization} We conclude by a visual comparison of the bibliographic coupling supernetwork using the same layout but different coloring according to its modularity class (Figure~\ref{fig:biblio_clustered_modularity}), ORES topics (Figure~\ref{fig:biblio_clustered_topics}), and WikiProjects (Figure~\ref{fig:biblio_clustered_wp}). These visualizations confirm two patterns we have previously highlighted: a) the important role of biographies in connecting this part of Wikipedia (first visualization); b) the systematic importance of STEM (second visualization). Furthermore, we can also appreciate the apparently effective role played by WikiProjects in helping editors coordinate for the curation of specialized knowledge in a coherent whole (third visualization).

\subsection*{Science from a Wikipedia perspective}

\begin{figure}[H]
\centering
\includegraphics[width=\textwidth]{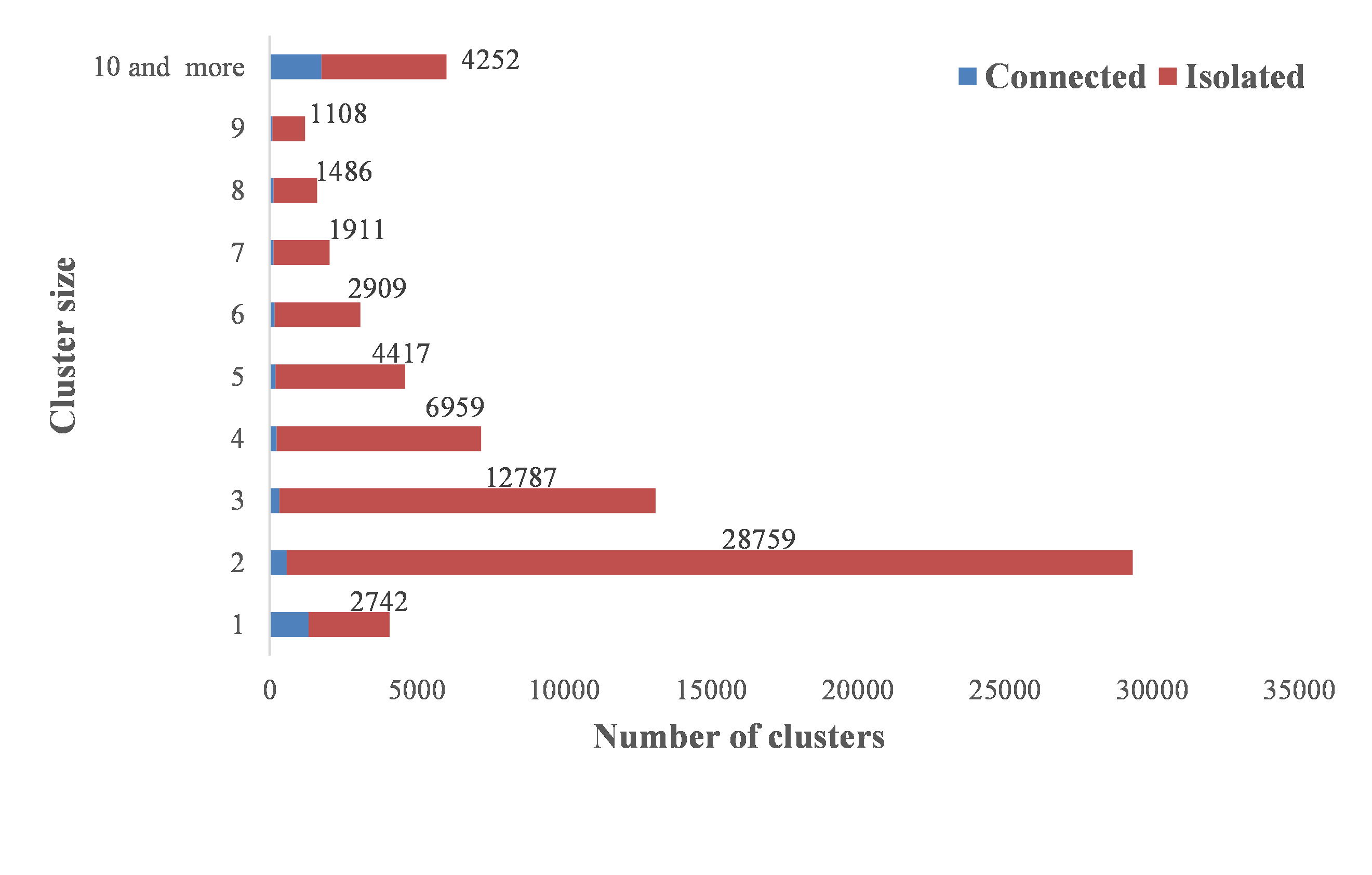}
\caption{Distribution of clusters by size in the co-citation network of journal articles.}
\label{fig:distribution_cocitation}
\end{figure}

\paragraph{Co-citation network} The co-citation network of journal articles cited from Wikipedia has 1,050,686 nodes and 17,916,861 edges, with a density of around 0.00003. The resulting supernetwork has 71,983 nodes (clusters) and 23,668 edges, and a density of 0.00001. The distribution of small to large cluster sizes and the fraction of isolated clusters is, consequently, even more pronounced than for the bibliographic coupling network, as it can be seen from Figure~\ref{fig:distribution_cocitation}.
  
     \begin{figure}[H]
    \centering
    \includegraphics[width=\textwidth]{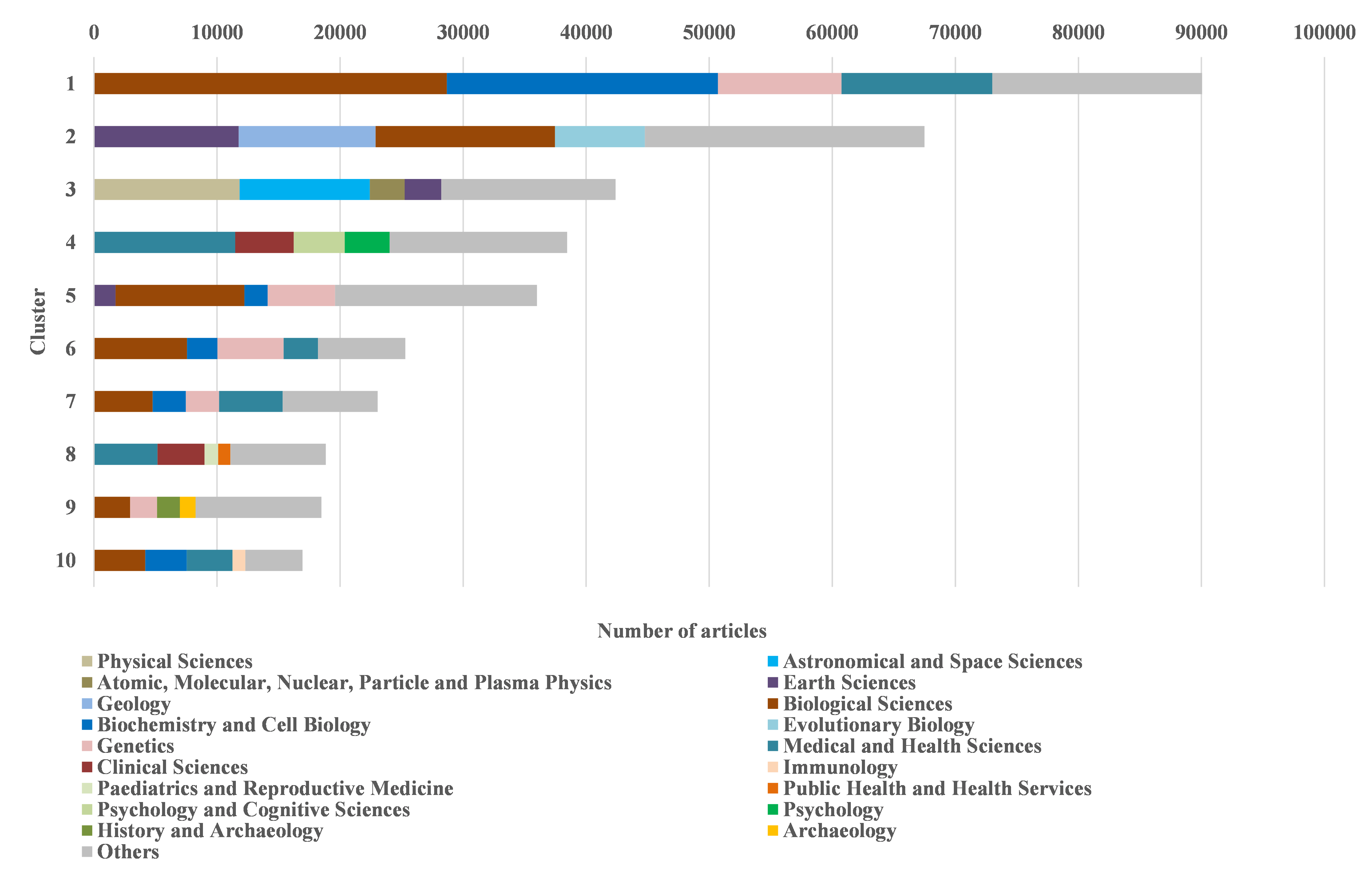}
    \caption{Distribution of major Fields of Research in the 10 largest clusters of the co-citation network of journal articles. The categories in each bar are ordered by color (in the legend), from left to right, top to bottom. For example, cluster one top major fields of research include (in order of appearance in the legend): Biological Sciences, Biochemistry and Cell Biology, Genetics, Medical and Health Sciences, Other.}
    \label{fig:cocit_top10}
    \end{figure}
    
\paragraph{Largest clusters} In Figure~\ref{fig:cocit_top10} we show the distribution of the top major Fields of Research per top-10 cluster in the co-citation network. This result strongly confirms the dominance of biology and medicine as the two top fields cited from Wikipedia. Virtually all top clusters include a dominant or at least sizable fraction of citations to these two fields, with the exception of the second cluster which contains a sizeable share of contributions in earth sciences, geology, and evolutionary biology, and the third one, focused on physics and astronomy.
    
\begin{figure}[H] 
\centering
\includegraphics[width=\textwidth]{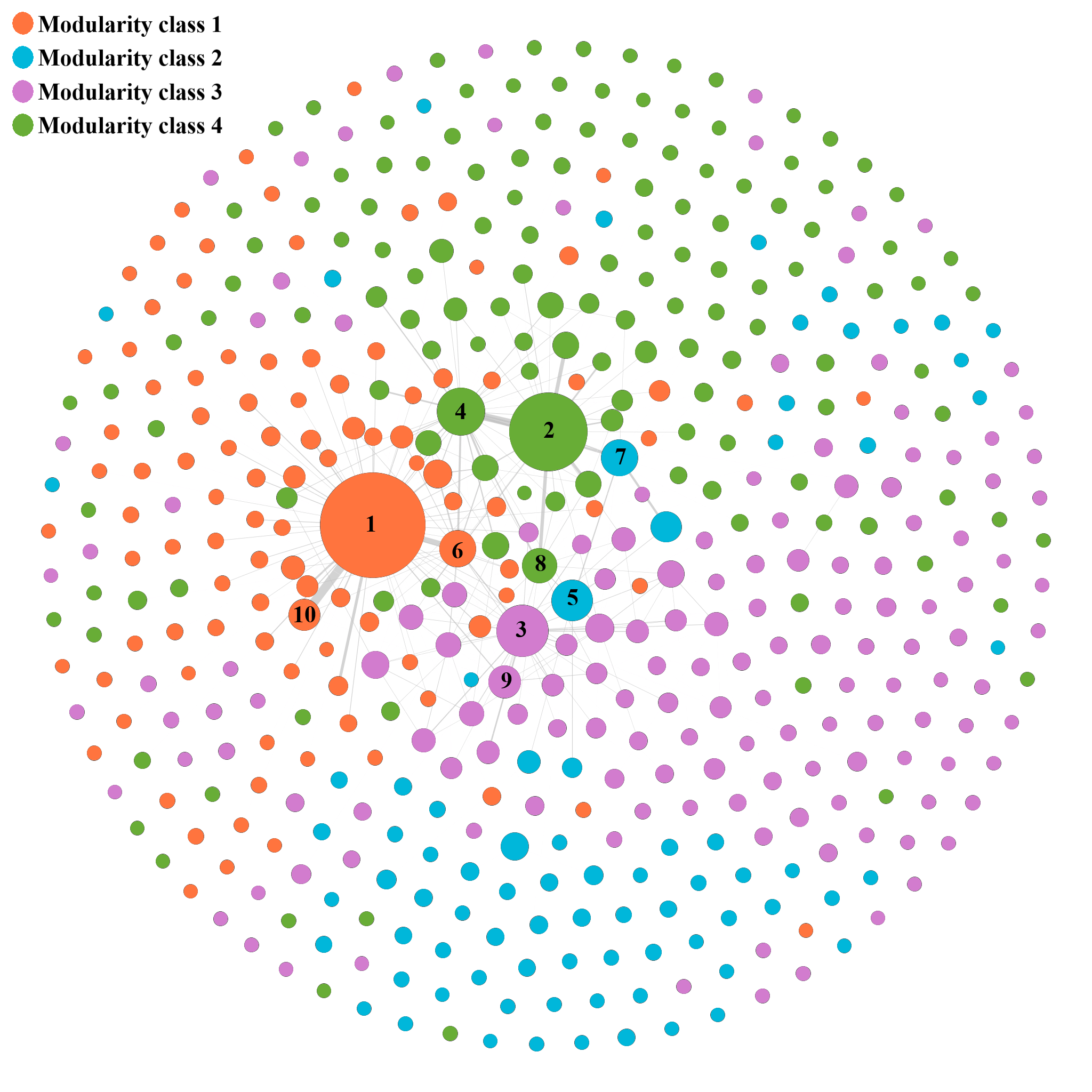}
\caption{Co-citation supernetwork coloured by modularity class.}
\label{fig:cocit_modularity}
\end{figure}

\begin{figure}[H]
\centering
\includegraphics[width=\textwidth]{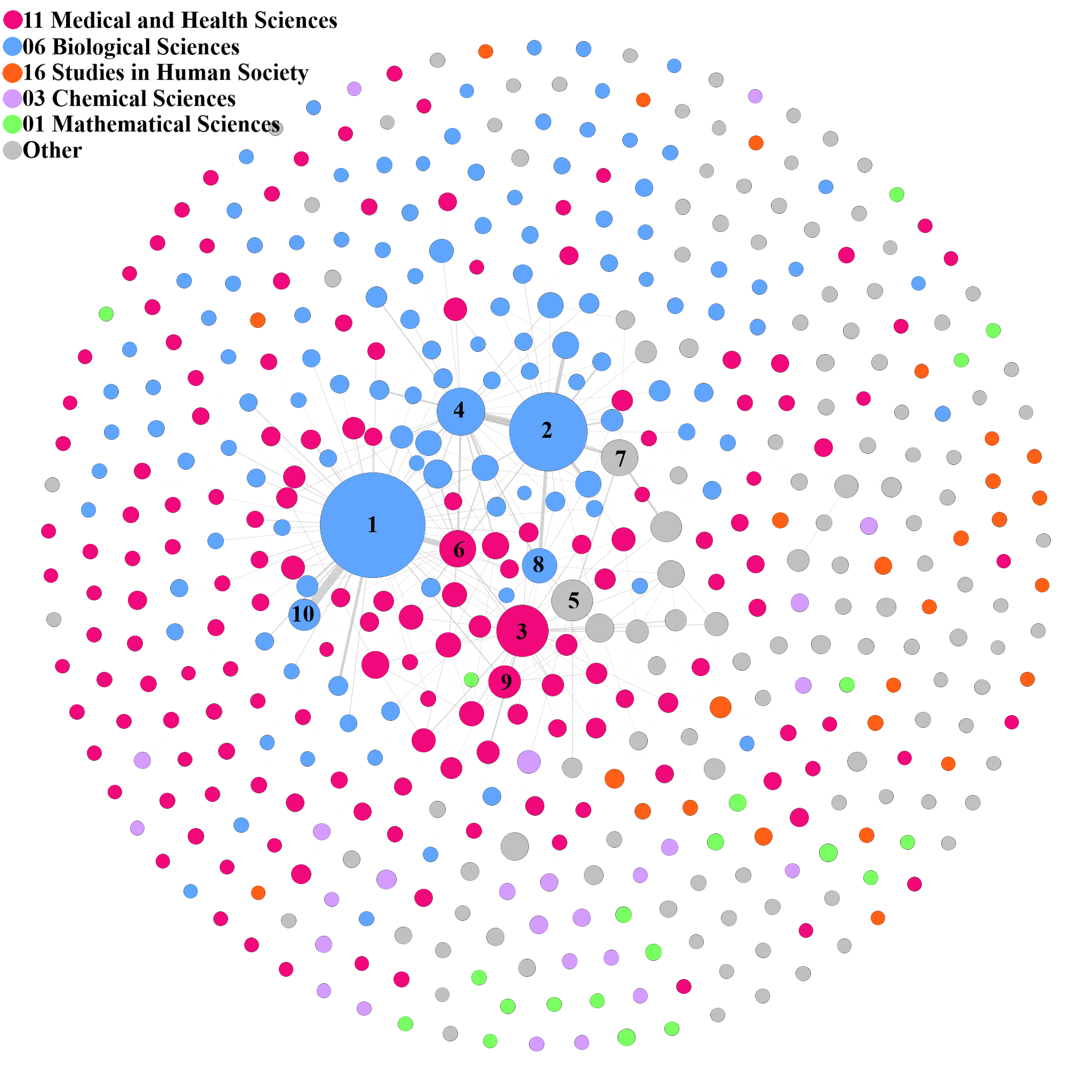}
\caption{Co-citation supernetwork coloured by top major Field of Research. Compare with Figure~\ref{fig:cocit_top10}.}
\label{fig:cocit_for}
\end{figure}

\paragraph{Network visualization} Also for the co-citation supernetwork we visualize it using the same layout and different coloring, respectively by modularity class (Figure~\ref{fig:cocit_modularity}) and top major Field of Research (Figure~\ref{fig:cocit_for}). The results are consistent with what we previously discussed. Furthermore, the two top clusters, which from Figure~\ref{fig:cocit_top10} we know being roughly focused on cell biology and genetics (cluster 1) and physical anthropology (cluster 2), group differently by modularity, with medicine, psychology and clinical sciences showing a stronger connection to the former than the latter.

\section*{Discussion}
    
We provided a map of science in Wikipedia highlighting clear trends, while rising some further questions. Firstly, our first research question asked which scientific sources are cited from Wikipedia and what are their characteristics. The dominant role of STEM literature clearly emerges, with biology and medicine as the top fields, followed by earth sciences, physics and astronomy. Furthermore, a sizeable fraction of this literature is well cited and published in notorious venues. These results confirm previous studies conducted on smaller datasets~\citep{colavizza_covid-19_2020} or at the journal level of analysis~\citep{arroyo2020science}. 

We were able to delve deeper as well, in view of our second research question asking which areas of Wikipedia rely on scientific sources and how do they relate. Firstly, the seemingly marginal role of journal articles from non-STEM fields is attenuated by the connecting role of biography articles in Wikipedia, which effectively bridge history, geography and culture with STEM topics. Secondly, the bibliographic coupling network of Wikipedia articles is not only smaller, but also better connected than the co-citation network of journal articles cited from Wikipedia. This might showcase the consolidating encyclopedic role of Wikipedia, as well as the positive impact of WikiProjects as a means to coordinate editorial efforts.

A set of new questions emerge from our work. To begin with, on the characteristics of journal articles cited from Wikipedia: we have found that on average these are well cited outside of Wikipedia too, yet many remain poorly cited or even not cited at all -- for example, 70\% of journal articles cited from Wikipedia received fewer than 100 citations from other journal articles at the time of the study. Furthermore, the fraction of Open Access articles cited from Wikipedia appears relatively high even though not dominant. Lastly, the age of such journal articles is mostly distributed within the past 20 years, possibly hinting at a chronological debt in Wikipedia: several citations might have been current when most of Wikipedia articles were created, but might no longer be up to date. Wikipedia is in this respect a `slow altmetric indicator', which takes time to accumulate~\citep{fang_studying_2020}. Secondly, our network analysis finds interesting clusters of articles which warrant further study. How the central role of the largest fields of research (biology, medicine) articulates with the networks' `periphery', in particular, remains an open question.

Some of the limitations of our study constitute possible directions for future work as well. The most important one is that we only considered citations to journal articles. Adding other cited sources, such as books and Web contents, would complement the map which we provided here. Secondly, the dataset we used constitutes a snapshot of Wikipedia at a certain point in time: a study of citations including time information would provide for a clearer picture of the dynamics of negotiation and consolidation of knowledge in Wikipedia. Lastly, Wikipedia's internal structure can also be mapped using information such as the internal link structure or the textual similarity of the articles. A comparison of the citations networks we studies here with these would further enrich the map of Wikipedia. 

\section*{Conclusion}

In this study, we mapped the organization of Wikipedia according to its use of scientific journal articles via citations. We made use of a recent dataset of citations from Wikipedia, and relied on network analysis techniques, in particular network clustering. We were able to show that Wikipedia heavily relies on scientific contents from biology, medicine and a handful of other STEM fields, including physics and earth sciences. Journal articles cited from Wikipedia are, on average, well-cited, published in notorious journals such as Nature or Science, and have been published over the past 20 years. While non-STEM fields are only marginally represented in journal articles cited from Wikipedia, they play an important connecting role via Wikipedia's biographies. This is but an example of how Wikipedia is able to interconnect knowledge across scientific fields and also with other non-scientific topics. In this respect, the most interesting future work which awaits us is the extension of this map of science in Wikipedia to include books, Web contents and all other sources cited in Wikipedia.

\subsection*{Code and data availability statement}

The code to replicate our work is made available online: \url{https://github.com/alsowbdxa/Code-of-Science-in-Wikipedia}. The Wikipedia Citations dataset is openly available~\citep{singh_wikipedia_2021}, while access to Dimensions can be requested through their portal. All other supporting datasets we used are openly available and referenced from the Data and Methods section.  

\subsection*{Acknowledgements}

We acknowledge the support of Dimensions in making their data available.

\theendnotes

\section*{Appendix}

\subsection*{Top-10 most cited journal articles}\label{sec:top_journal_articles}

\begin{enumerate}
    \item Laemmli U. K. (1970) ``Cleavage of Structural Proteins during the Assembly of the Head of Bacteriophage T4''. 
    \textit{Nature}.
    \textbf{Open access: closed}.
    \textbf{Research organization country: United Kingdom}.
    \textbf{Number of citations: 214,886}. 
    \textbf{Number of recent citations: 6,111}.
    
    \item Bradford M. M. (1976) ``A rapid and sensitive method for the quantitation of microgram quantities of protein utilizing the principle of protein-dye binding''. \textit{Analytical Biochemistry}.
    \textbf{Open access: closed}.
    \textbf{Research organization country: United States}.
    \textbf{Number of citations: 193,330}. 
    \textbf{Number of recent citations: 16,578}.
    
    \item Perdew J. P., Burke K., Ernzerhof M. (1996) ``Generalized Gradient Approximation Made Simple''. \textit{Physical Review Letters}.
    \textbf{Open access: closed}.
    \textbf{Research organization country: United States}.
    \textbf{Number of citations: 99,164}. 
    \textbf{Number of recent citations: 27,949}.
    
    \item G.M. Sheldrick. (2007)  ``A short history of SHELX''. \textit{ Acta Crystallographica Section A: Foundations and advances}.
    \textbf{Open access: open}.
    \textbf{Research organization country: Germany}.
    \textbf{Number of citations: 72,560}. 
    \textbf{Number of recent citations: 7,355}.
    
    \item Axel D. Becke. (1993)  ``Density‐functional thermochemistry. III. The role of exact exchange''. \textit{The Journal of Chemical Physics}.
    \textbf{Open access: closed}.
    \textbf{Research organization country: Canada}.
    \textbf{Number of citations: 69,187}. 
    \textbf{Number of recent citations: 8,871}.
    
    \item Chengteh Lee, Weitao Yang, Robert G. Parr. (1988) ``Development of the Colle-Salvetti correlation-energy formula into a functional of the electron density''. \textit{Physical Review B}.
    \textbf{Open access: closed}.
    \textbf{Research organization country: United States}.
    \textbf{Number of citations: 66,421}. 
    \textbf{Number of recent citations: 8,872}.
    
    \item Marshal F. Folstein, Susan E. Folstein, Paul R. McHugh. (1975)  ``“Mini-mental state” A practical method for grading the cognitive state of patients for the clinician''. \textit{Journal of Psychiatric Research}.
    \textbf{Open access: closed}.
    \textbf{Research organization country: United States}.
    \textbf{Number of citations: 64,625}. 
    \textbf{Number of recent citations: 8,428}.
    
    \item Stephen F. Altschul, Warren Gish, Webb Miller, Eugene W. Myers, David J. Lipman. (1990) ``Basic local alignment search tool''. \textit{Journal of Molecular Biology}.
    \textbf{Open access: closed}.
    \textbf{Research organization country: United States}.
    \textbf{Number of citations: 63,340}. 
    \textbf{Number of recent citations: 10,717}.
    
    \item F. Sanger, S. Nicklen, A. R. Coulson. (1977) ``DNA sequencing with chain-terminating inhibitors''. \textit{PNAS}.
    \textbf{Open access: open}.
    \textbf{Research organization country: United Kingdom}.
    \textbf{Number of citations: 58,637}. 
    \textbf{Number of recent citations: 1,046}.
    
    \item Piotr Chomczynski, Nicoletta Sacchi. (1987) ``Single-step method of RNA isolation by acid guanidinium thiocyanate-phenol-chloroform extraction''. \textit{Analytical Biochemistry}.
    \textbf{Open access: closed}.
    \textbf{Research organization country: United States}.
    \textbf{Number of citations: 56,286}. 
    \textbf{Number of recent citations: 993}.
    
\end{enumerate}

\subsection*{Figures}\label{sec:appendix figures}

\bibliography{bibliography}

\end{document}